# Non-linear field theory III.
## Geometrical illustration of the electromagnetic representation of Dirac's electron theory

Alexander G. Kyriakos

*Saint-Petersburg State Institute of Technology,
St. Petersburg, Russia*

*Present address: Athens, Greece, e-mail: agkyriak@yahoo.com*

**Abstract**

The present paper is the continuity of the previous papers "Non-linear field theory" I and II. Here on the basis of the electromagnetic representation of Dirac's electron theory we consider the geometrical distribution of the electromagnetic fields of the electron-positron and on this base give the explanation and solution of many fundamental problems of the QED

PASC 12.10.-g  Unified field theories and models.
PASC 12.90.+b  Miscellaneous theoretical ideas and model.

## 1.0. Introduction

The present paper is the continuity of the previous papers [1,2]. Here on the basis of the electromagnetic representation of Dirac's electron theory the results, obtained in previous papers, are geometrically illustrated and completed.
    As it is known in the framework of the quantum electrodynamics the photon and electron (positron) do not have a structure. In other words they are the point particles and has not the geometrical characteristics.
    On the other hand, the mathematical description of an electromagnetic field both of a photon and of an electron by vectors and tensors means an opportunity of geometrical representation of some characteristics of these particles. For example: the arrangement of vectors in space relatively to each other (parallelism, perpendicularity, etc.), the trajectories of their movement (rectilinear and curvilinear) and their arrangement in space, the way of their transport (parallel transport of vectors, their rotation around of some point, etc.), the symmetry of the vector field (spherical, cylindrical, etc.), transformations and preservation of vectors concerning various

transformations of space (transformations of rotation, shift, etc.) - all this allows to create evident geometrical images of these objects.

So since the electron has some space distribution of the electromagnetic fields, we can talk about geometrical characteristics of this electromagnetic field distribution.

Below on the base of the geometrical characteristics of the electromagnetic field space distribution we give the explanation and solution of many fundamental problems of the QED. (Note, that these results will be written in *italic*).

## 2.0. Geometrical characteristics of the electromagnetic wave and photon

Let us state briefly the data, which the physics has accumulated for today, concerning a photon.

A photon is the particle or quantum of an electromagnetic wave, which contains the mutually perpendicular magnetic and electric fields, changing in time according to the harmonic law and having the certain polarization.

In accordance with the Planck - Einstein theory [3], a monochromatic EM wave will consist of $N$ monoenergetic photons, each of which has the following energy $\varepsilon_p$ and momentum $\vec{p}_p$:

$$\varepsilon_p = h\nu = \hbar\omega, \quad \vec{p}_p = \hbar\vec{k}$$

where $\nu$ and $\omega$ are the linear and circular frequencies accordingly, $h$ and $\hbar$ are the usual and "bar" Planck's constants.

The photon as a corpuscle has no rest mass, since it moves with the speed of the light $c$. The number of the photons in an electromagnetic wave is such, that their full energy $W = N \cdot \varepsilon = N\hbar\omega$ is equal to the full EM energy of a wave. The angular moment (or spin) of one photon is equal to $J_p = \hbar$. It is also known, that in accordance with the principle of the particles identity, all photons of identical energy and polarization are completely identical. Numerous experiments prove the localness of the photon (Vavilov's experiments, etc.) [4].

Within the framework of quantum electrodynamics the sizes of a photon cannot be defined and it is impossible to tell anything about its structure. Actually, in QED [5] the wave function in the coordinate representation $f(\vec{r},t)$ is meaningless, as it is not defined by the value of the electromagnetic field $\vec{E}(\vec{r},t)$ in the same point, but **depends on the distribution of the field in some area, whose sizes are about the wavelength**. The last means, that the localization of the photon in a smaller area is impossible. However this conclusion testifies indirectly the presence of some volume and structure at the photon.

### 2.1. Lorentz-invariant analysis

Let us try to make some meaning about a photon model.

The above experimental data and theory form the basis of the following statement: *all photons of identical energy and polarization are identical on structure and contain one period of a wave.* Actually, if the same photon would be able to have a different number of wavelengths and different structure, it would be difficult to explain the existence of the Planck formula and the principle of identity.

Since a photon is characterized only by one frequency, we must state that *any photon is monochromatic (i.e. one-frequency or mono-energetic) particle*.

Let us consider now a coherent monochromatic light beam of the round cross-section with the area $S_{beam}$, which has some longitudinal size $l_{beam}$. Such beam can be named a zug or a wave packet. Experimentally it can be received from the laser. Since the photons are bosons, they have the possibility at the identical energy to merge so, that all of them move in the same phase and build the monochromatic electromagnetic wave. Thus, it is possible to consider a laser zug as a piece of a monochromatic electromagnetic wave of the certain frequency, amplitude and energy.

Obviously, the zug has an integer number of wavelengths. Hence, it can be reduced, by rejecting pieces, multiple to the number of waves. In the experiment, zugs with a length, which is equal to some wavelengths, have already been received. In 1983 [6] in the Bell Labs, laser pulses in 30 fmc were received, which corresponds approximately to 14 periods of a light wave. Now pulses which have duration of 4 periods of a wavelength [6], are received. Theoretically there is no limit in receiving a zug of one wavelength. Thus, it is necessary to assume the zug of one wavelength, as the least part of an electromagnetic wave. Since the least part of an electromagnetic wave is the photon, hence, its length in the direction of the wave propagation is equal to one wavelength.

Which limit of diameter of a beam do we obtain, if we reduce the zug cross-section? The experimental limit is one wavelength, as then become essentially the diffraction phenomena. Theoretically the lower limit of the diameter of a beam should be limited to a photon cross-section size. But the theory does not speak, whether it is less than one wavelength. Nevertheless, if to consider, that all photons of the same energy are identical, the photon should have the certain cross-section size.

The essential data on the photon characteristics is given by the invariant analysis of an electromagnetic field. As is known, [7] (page 30-31) for the zug of a plane electromagnetic wave the following relations are Lorentz-invariant ($L-inv$):

$$E_0/\omega = E_0'/\omega' \sim L-inv$$
$$\varepsilon/\omega = \varepsilon'/\omega' \sim L-inv \quad (2.1)$$
$$V \cdot \omega = V' \cdot \omega' \sim L-inv$$

where $V$, $E_0$, $\varepsilon$, $\omega$ are the volume, amplitude, energy and cyclic frequency of a wave packet correspondingly.

If we supposed, that this invariance take also place for the characteristics of separate photons, for a photon of any frequency the following relations are fair:

$$E_0/\omega = const_1 = C_1$$
$$\varepsilon/\omega = const_2 = C_2 \quad (2.2)$$
$$V \cdot \omega = const_3 = C_3$$

(Actually, choosing a speed of the primed inertial system, we will register the various frequencies $\omega'$ of the same photon, and for all of them, the relations (2.1) will be observed).

From here follows, that for a photon of a certain frequency, the certain

amplitude: $\quad E_0 = C_1 \cdot \omega$, (2.3)

energy: $\quad \varepsilon = C_2 \cdot \omega$, (2.4)

and volume: $\quad V = C_3/\omega$, (2.5)

are available.

Comparing (2.3) with the Planck's formula, it is not difficult to define the constant $C_2$:

$$C_2 = \hbar$$

(Note that the fact that the Planck formula follows from the analysis of the Lorentz-invariance of the electromagnetic theory, was find long time ago [8]).

The model of a photon should satisfy one more relation: as is known, the photon has angular momentum equal to $1\hbar$. Thus, *the photon field should rotate so that his angular momentum is equal to $1\hbar$, and the circular frequency of rotation of a photon fields is equal to the photon wave frequency.*

## 2.1 A heuristic photon model and its characteristics

We have no experimental data about the geometrical form of a photon. Proceeding from reasons of symmetry, three forms of a photon are generally possible: cylindrical, spherical and toroidal. But it is difficult to connect linear propagation of an electromagnetic wave with the spherical form; the cylindrical form demands cutting a wave, that breaks monochromaticy (mono-frequency) of a photon. Thus, it remains only a toroidal form, but the analysis of such form of a photon is inconvenient and demands the separate research.

Therefore we will assume, that as a first approximation the photon has the cylindrical form, and we shall try to define characteristics of such photon, proceeding from requirements of the our theory.

Let us analyze the formula (2.5). Assuming, that the photon represents the cylinder of the length $\lambda$ and some cross-section $S_p$, we will receive for its volume size:

$$V = S_p \cdot \lambda , \qquad (2.6)$$

But from the formula (2.5), taking into account, that $\omega = \dfrac{2\pi c}{\lambda}$, it is no difficult to receive:

$$V = \frac{C_3}{2\pi c}\lambda , \qquad (2.7)$$

Comparing (2.6) and (2.7), we come to the conclusion, that:

$$S_p = \frac{C_3}{2\pi c} = const , \qquad (2.8)$$

Thus, in first approximation a photon can be imagined as the classical relativistic strings of certain length and cross-section, oscillated with the certain frequency. So as heuristic model of a photon we can accept some volume of EM excitation, which has the form of the circular cylinder, whose radius and length have the order of the wavelength of the photon – see fig. 2.1:

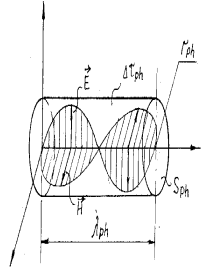

Рис. 2.1.

where $\lambda_p$, $r_p$, $S_p$, $\Delta\tau_p$ are accordingly the wavelength of a photon, its cross-section radius, the area of cross-section and the volume (conditionally inside the volume of the excitation, the electric and magnetic fields distribution $\vec{E}$, $\vec{H}$ are represented).

Since we consider a photon as the source that produces the electron-positron pair, the above-stated characteristics are defined by the pair production conditions. In particular, this photon should have: the energy $\varepsilon_p = 2m_e c^2$ (where $m_e$ is the electron mass) and the circular frequency $\omega_p = \varepsilon_p / \hbar = 2m_e c^2 / \hbar$. Thus it is not difficult to calculate the wavelength of such photon:

$$\lambda_p = \frac{2\pi c}{\omega_p} = \frac{\pi \hbar}{m_e c}.$$

According to the accepted model $\lambda_p = 2\pi r_p$. Then $r_p = \frac{\hbar}{2m_e c}$, $S_p = \pi r_p^2 = \frac{\pi}{4}\left(\frac{\hbar}{m_e c}\right)^2$, $\Delta\tau_p = \lambda_p \cdot S_p = 2\pi^2 r_p^3$. (It is interesting that the last formula corresponds exactly to the volume of the torus, whose both radii are equal to $r_p = \frac{\hbar}{2m_e c}$).

About the minimal length of a photon of the known energy it is also possible to make a conclusion on the basis of the Heisenberg uncertainty principle [5,9]. Actually, the product of the length of quantum object on his momentum should correspond to the relation:

$$\lambda \cdot p_p \geq 2\pi \hbar, \tag{2.9}$$

whence:

$$\lambda \geq \frac{2\pi \hbar}{p_p} = 2\pi \frac{\hbar c}{\varepsilon_p} = \frac{\pi \hbar}{m_e c} = 2\pi \frac{1}{\alpha} \frac{e^2}{\varepsilon_p}, \tag{2.10}$$

that coincides with the above-stated values (here $\alpha$ is the thin structure constant, and $e$ is the electron charge). About the cross-section size of a photon, it is impossible to tell anything from the uncertainty relation point of view. (Note that only the toroidal model of a photon gives this value of the angular momentum; actually, putting the photon "mass" $m_p = 2m_e$, for the torus angular momentum we have: $J = m_p r_p^2 \omega_p = \hbar$)

## 3.0. Geometrical illustration of the particle-antiparticle production process

Let's consider step by step the particle-antiparticle pair production conditions.
$$\gamma + Ze \rightarrow e^+ + e^- + Ze, \qquad (3.1)$$
where $\gamma$, $e^+$, $e^-$ are the photon, positron and electron correspondingly, $Ze$ means the electromagnetic field of the other particle (e.g. of the nucleus).

### 3.1. The interaction of the photon with electromagnetic field of the nucleus

In the previous paper [1] we have showed that *by the fulfilment of the pair production conditions the electromagnetic wave (photon) is able to move along the closed curvilinear trajectory, making some stable construction named elementary particle; as a result, the transformation of the electromagnetic field of the photon can take place and the linear electromagnetic wave, characterised by field vectors, transforms into the circular electromagnetic wave, characterised as field spinors.*

This transformation could be illustrated by following scheme of the electron-positron pair production process:

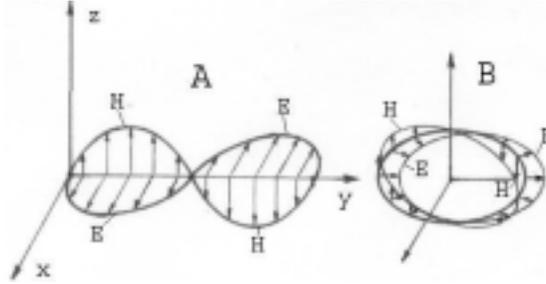

Fig. 1

We can suppose that *the origin of the electromagnetic wave twirling is the bending of the electromagnetic wave in the electromagnetic field of the nucleus as in the medium with high refraction index.*

In connection with our representations, a question arises: which equation describes the twirled photon B (fig. 1)? It is not difficult to understand, that this equation is the Klein-Gordon equation [5,9,10]:
$$\left(\hat{\varepsilon}^2 - c^2 \hat{\vec{p}}^2 + m_p^2 c^4\right)\psi = 0, \qquad (3.2)$$
where $\hat{\varepsilon} = i\hbar \dfrac{\partial}{\partial t}, \hat{\vec{p}} = -i\hbar \vec{\nabla}$ are the operators of the energy and momentum, $c$ is the light velocity, $m_p$ is the twirled photon mass. It is also the wave equation [1,2] with the particle mass term:
$$\left(\nabla^2 - \frac{1}{c^2}\frac{\partial^2}{\partial t^2}\right)\psi = \frac{m_p^2 c^2}{\hbar^2}\psi, \qquad (3.3)$$

The equation (3.2) can also be written in the following form:

$$\left[ \left( \hat{\alpha}_o \hat{\varepsilon} \right)^2 - c^2 \left( \hat{\vec{\alpha}} \cdot \hat{\vec{p}} \right)^2 + \left( \hat{\beta} m_p c^2 \right)^2 \right] \psi = 0, \qquad (3.4)$$

where $\hat{\alpha}_o = \hat{1}$, $\hat{\vec{\alpha}}$, $\hat{\alpha}_4 \equiv \hat{\beta}$ are the Dirac matrices. Actually, the equation (3.4) can be disintegrated on two linear equations of the electron and positron. In other words *it is the equation of the electron-positron bounded state, i.e. the equation of the twirled photon.*

But it is known [10], that the Klein-Gordon equation is scalar, since his wave function has only one component. Such equation describes a particle with zero spin. But it appeared that this equation cannot be considered as the quantum equation, since according to this equation, the probability of detection of a particle in the certain point of space can accept negative values, which is senseless.

The correct interpretation of the Klein-Gordon equation was given by W. Pauli and V. Weiskopf [11]. They suggest to consider this equation as the classical equation of the field (similarly to the equations of an electromagnetic field) and then it to quantize. In this case the Klein-Gordon equation describes a particle with any spin, since it correspond to the energy–momentum conservation law.

Another feature of the equation of Klein-Gordon consists in his connection with the Dirac equation. According to our representations, the Dirac equation is the Maxwell equations for twirled electromagnetic waves while the Klein-Gordon equation is the wave equation for the twirled waves. As H. Bethe [10] has noted, a similar situation takes place in electrodynamics, but to this circumstance, neither he, nor the others have given any weight. The Maxwell equations are the equations of the first order with regard to all variables. On the other hand, every field component satisfies the wave equation of the second order, which is similar to the Klein - Gordon equation, but with zero rest mass. These two requirements do not contradict with each other due to the fact that each Maxwell equation connects various components of the field.

Thus, the Klein-Gordon equation, containing the spinor (or bispinor) as wave function, describes the twirled photon (or two photons of one direction in the case of the bispinor), i.e. the particle with unit spin.

Now we will consider in detail the geometric representation of the electromagnetic fields transformation in the electron-positron pair production process.

### 3.2. The electromagnetic wave twirling and the mass-current term appearance

Let the plane-polarized wave, which have the field vectors $(E_x, H_z)$ (see fig. 2)

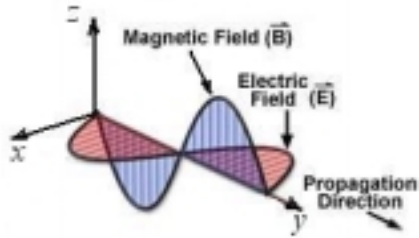

Fig. 2

be twirled with some radius $r_K$ in the plane $(X',O',Y')$ of a fixed co-ordinate system $(X',Y',Z',O')$ so that $E_x$ is parallel to the plane $(X',O',Y')$ and $H_z$ is perpendicular to it (fig. 3).

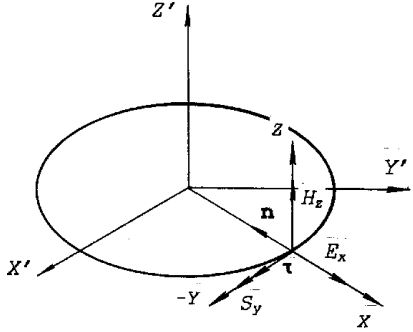

Fig. 3

According to Maxwell [12] the displacement current is defined by the equation:

$$j_{dis} = \frac{1}{4\pi}\frac{\partial \vec{E}}{\partial t}, \qquad (3.5)$$

The above electrical field vector $\vec{E}$, which moves along the curvilinear trajectory (let it have the direction from the centre), can be written in the form (see fig. 4):

$$\vec{E} = -E \cdot \vec{n}, \qquad (3.6)$$

where $E = |\vec{E}|$, and $\vec{n}$ is the normal unit-vector of the curve (having direction to the center).

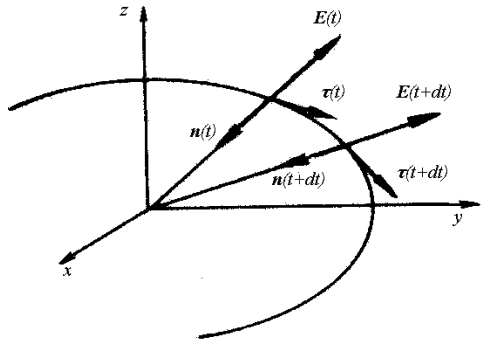

Fig. 4

According to the differential geometry the derivative of $\vec{E}$ with respect to $t$ can be represented as:

$$\frac{\partial \vec{E}}{\partial t} = -\frac{\partial E}{\partial t}\vec{n} - E\frac{\partial \vec{n}}{\partial t}, \qquad (3.7)$$

Here the first term has the same direction as $\vec{E}$. The existence of the second term shows that at the twirling of the wave the supplementary displacement current appears. It is not difficult to show that it has a direction, tangential to the ring:

$$\frac{\partial \vec{n}}{\partial t} = -v_p K \vec{\tau}, \qquad (3.8)$$

where $\vec{\tau}$ is the tangential unit-vector, $v_p \equiv c$ is the electromagnetic wave velocity, $K = \dfrac{1}{r_K}$ is the curvature of the trajectory and $r_K$ is the curvature radius. Thus, the displacement current of the ring wave can be written in the form:

$$\vec{j}_{dis} = -\frac{1}{4\pi}\frac{\partial E}{\partial t}\vec{n} + \frac{1}{4\pi}\omega_K E \cdot \vec{\tau}, \qquad (3.9)$$

where $\omega_K = \dfrac{v_p}{r_K} \equiv cK$ we name the curvature angular velocity, $\vec{j}_n = \dfrac{1}{4\pi}\dfrac{\partial E}{\partial t}\vec{n}$ and $\vec{j}_\tau = \dfrac{\omega_K}{4\pi}E \cdot \vec{\tau}$ are the normal and tangent components of the current of the twirled electromagnetic wave, correspondingly. Thus

$$\vec{j}_{dis} = \vec{j}_n + \vec{j}_\tau, \qquad (3.10)$$

From full theory followed (see section of the paper) that the mass term of the Dirac equation in the electromagnetic representation is equivalent to the following electric current:

$$\vec{j}^e = i\frac{\omega}{4\pi}\vec{E} = i\frac{c}{4\pi}\frac{1}{r_C}\vec{E}, \qquad (3.11)$$

where $\omega = \dfrac{mc^2}{\hbar}$, and $r_C = \dfrac{\hbar}{mc}$ is the Compton length wave of the electron.

As we see the current $\vec{j}_\tau = \dfrac{\omega_K}{4\pi}E \cdot \vec{\tau}$ corresponds to the electrical current (2.22), if we taken into account that the imaginary unit correspondents to the rotation on $\pi/2$ and $r_K = r_C$. (Note that it can show that when the electromagnetic wave has the circular polarization the magnetic current (2.23) appears also, but integrally this current is equal to zero).

The currents $\vec{j}_n$ and $\vec{j}_\tau$ are always mutually perpendicular, so that we can write in complex form:

$$j_{dis} = j_n + ij_\tau, \qquad (3.12)$$

where $i = \sqrt{-1}$.

Thus, *the origin of the appearance of the imaginary unity in the quantum mechanics is tied with the tangent current, that explain why the Dirac theory is the complex theory.*

## 4.0. Electromagnetic model of electron

According to our results [1,2] the electron (positron) is «created» from the curvilinear electromagnetic wave fields, which is described by quantum electrodynamics. In other words this means that the electron (positron) does not have some "core", i.e. they are represented as the point particles. These results are verified from the results of the quantum electrodynamics. According to the modern experiments the quantum electrodynamics is correct until 2x10$^{-18}$ m). Actually, *if the quantum electrodynamics is the electrodynamics of curvilinear (i.e. non-linear) waves [1,2], the quantum electrodynamics must be correct on any short distance.*

## 4.1. Structure and parameters of the electron model

According to modern representations the physical field represents a continuum. Hence, the electron, as exclusively electromagnetic particle, represents some space volume, continuously occupied with an electromagnetic field.

As we have shown, in the electromagnetic representation of the Dirac electron theory this volume has a toroidal symmetry and some spatial (geometrical) characteristics.

Such representation is proved by the analysis of the Dirac equation. So it is known, that [13] "in the non-relativistic limit the electron does not represent a point charge, but the distribution of charge and current in the area with linear sizes $\frac{\hbar}{m_e c}$. This explains the occurrence of interaction terms, which are connected to the magnetic moment (interaction $\vec{\mu}\vec{H}$, spin-orbital interaction) and the distributed density of charges (Darvin's component)".

In classical electrodynamics of Lorentz the electron energy consists exclusively of its electromagnetic field. Here it is shown, that if we consider an electron field as a sphere, it is possible to calculate its radius, which characterizes the volume of the electromagnetic field, including the basic part of the electron energy. It is called classical electron radius and is equal to $r_o = \frac{e^2}{mc^2}$.

Similarly to the classical radius we will enter the sizes of the toroidal volume including the basic part of the electron energy (fig. 5):

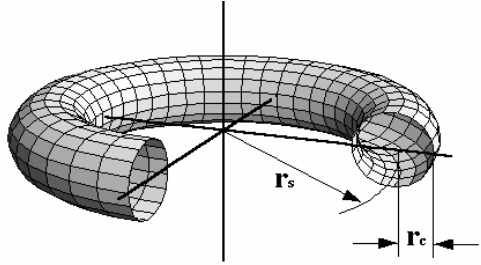

Fig. 5

It is obvious that this torus radius is equal to $r_s = \frac{\lambda_p}{2\pi}$, where $\lambda_p$ is the photon's wavelength.

In our case the photon characteristics are defined by the electron-positron pair production conditions: namely by the photon energy $\varepsilon_p = 2m_e c^2$ and by the circular frequency $\omega_p = \frac{\varepsilon_p}{\hbar} = \frac{2m_e c^2}{\hbar}$. Therefore, the photon wavelength is $\lambda_p = \frac{2\pi c}{\omega_p} = \frac{\pi \hbar}{m_e c}$. Taking in account this value we can write: $r_s = \frac{\hbar}{2m_e c} = \frac{1}{\alpha}\frac{e^2}{2m_e c^2}$. We can suppose that in the general case the cross-section torus radius $r_c$ (fig. 5) is equal to $r_c = \zeta\, r_s$, where $\zeta \leq 1$.

Since the Dirac equation includes only one torus characteristic - the Compton radius, it is possible to assume, that both radiuses are identical (i.e. $\zeta = 1$) and the torus has the following form (fig. 6):

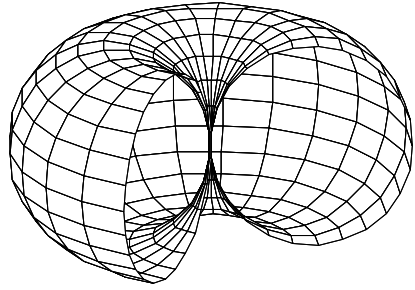

Fig. 6

## 4.2. Charge appearance and division hypothesis

Let us consider how the division of the twirled photon $B$ (fig. 1) takes place?
It is not difficult to calculate the charge density of the twirled photon:

$$\rho_p = \frac{j_\tau}{c} = \frac{1}{4\pi}\frac{\omega_p}{c}E = \frac{1}{4\pi}\frac{1}{r_p}E, \qquad (4.1)$$

The full charge of the twirled photon can be defined by integrating along all the torus volume $\Delta\tau_t$:

$$q = \int_{\Delta\tau_t} \rho_p d\tau, \qquad (4.2)$$

Using the model (fig. 4) and taking $\vec{E} = \vec{E}(l)$, where $l$ is the length of the way, we obtain:

$$q = \int_{S_t}\int_0^{\lambda_p} \frac{1}{4\pi}\frac{\omega_p}{c}E_o \cos k_p l \, dl \, ds = \frac{1}{4\pi}\frac{\omega_p}{c}E_o S_c \int_0^{\lambda_p} \cos k_p l \, dl = 0, \qquad (4.3)$$

(here $E_o$ is the amplitude of the twirled photon wave field, $S_c$ - the area of torus cross-section, $ds$ is the element of the surface, $dl$ - the element of the length, $k_p = \frac{\omega_p}{c}$ - the wave-vector).

It is easy to understand these results: because the ring current changes its direction every half-period, the full charge is equal to zero. Therefore, *the above model may represent only the non-charged particles*.

It is clear that a charge particle must contain only one half-period of wave. Thus *at the moment when the photon begins to roll up, the spontaneous photon division in two half-periods must take place* (fig. 7).

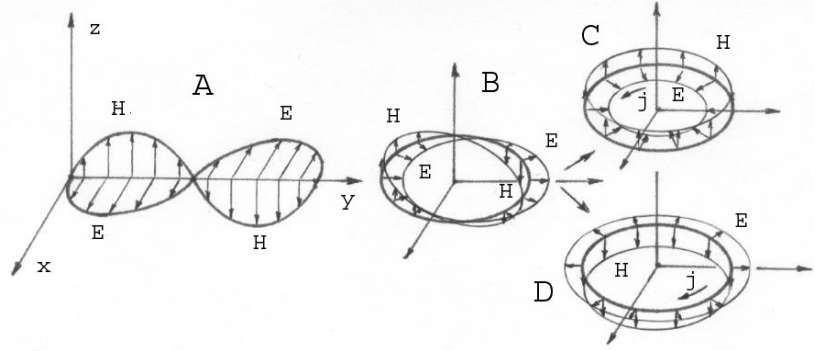

Fig. 7

It is clear that here (fig. 7) the parts *C* and *D* (two twirled semi-photons) contain the currents of opposite directions. Thus we can suppose, that *the cause of a twirled photon division is the mutual repulsion of oppositely directed currents.*

It is also clear that *the division process corresponds to the process of the particle-antiparticle pair production.* It's proved out by the fact that *the "daughters" twirled semi-photons C and D are completely anti-symmetric and can't be transformed one to each other by any transformation of co-ordinates (if it is not to be accompanied by the changing of field direction).*

It is interesting that *both the twirled semi-photon and the twirled photon radii must be the same.* This fact follows from the angular momentum conservation low. Really, the twirled photon angular momentum is equal to:

$$\sigma_p = p_p \cdot r_p = 2m_e c \cdot \frac{\hbar}{2m_e c} = 1\hbar, \tag{4.4}$$

In accordance with the momentum conservation low:

$$\sigma_s^+ + \sigma_s^- = \sigma_p, \tag{4.5}$$

where $\sigma_s^+, \sigma_s^-$ are the spins of the plus and minus semi-photons (i.e. of the electron and positron fields). Then we obtain:

$$\sigma_s = \frac{1}{2}\sigma_h = \frac{1}{2}\hbar, \tag{4.6}$$

Since

$$\sigma_s = p_s \cdot r_s, \tag{4.7}$$

where $r_s$ is the twirled semi-photon (electron) radius, and $p_s = m_e c$ is the inner semi-photon (electron) momentum, we have

$$r_s = \frac{\sigma_s}{p_s} = \frac{1}{2}\frac{\hbar}{m_e c} = \frac{\hbar}{2m_e c} = r_p, \tag{4.8}$$

Thus, *the torus size of the twirled photon doesn't change after division.*

As it is not difficult calculate, *the angular velocity (angular frequency) also doesn't change*: $\omega_s = \frac{c}{r_s} = \frac{2m_e c^2}{\hbar} = \omega_p$. From this it also follows that *the volumes and areas of the twirled photon and the twirled semi-photon will be the same*: $\Delta \tau_s = \Delta \tau_p$, $S_s = S_c$

The division of the twirled photon makes it possible to outline the solution of the some fundamental problems:

1) *the origin of the charge conservation low*: since in nature there are the same numbers of the photon half periods, the sum of the particles charge is equal zero.

2) *the difference between positive and negative charges*: this difference follows from the difference of the field and tangent current direction of the twirled semi-photons after pair production (by condition that the Pauli - principle is true).

3) *Zitterbewegung*. The results obtained by E. Schroedinger in his well-known articles about the relativistic electron [14] are the most important confirmation for the electron structure model. He showed, that electron has a special inner motion "Zitterbewegung", which has frequency $\omega_z = \frac{2m_e c^2}{\hbar}$, amplitude $r_z = \frac{\hbar}{2m_e c}$, and velocity of light $v = c$. The attempts to explain this motion had not given results. But if the electron is the twirled semi-photon, then we receive the simple explanation of Schroedinger's analysis.

4) *the difference between bosons and fermions* is well explained: the bosons contain the even number and the fermions contain the odd number of the twirled semi-photons.

5) *infinite problems*: in the non-linear theory the charge and mass infinite problems don't exist.

## 5.0. The characteristics of the electron-positron.

### 5.1. Charge and mass of twirled semi-photon

We can now calculate the twirled semi-photon charge:

$$q_s = \frac{1}{\pi}\frac{\omega_s}{c}E_o S_s 2\int_0^{\frac{\lambda_s}{4}} \cos k_s l \, dl = \frac{1}{\pi}E_o S_c, \tag{5.1}$$

Since $S_s = \pi r_c^2$, we obtain:

$$q_s = E_o r_c^2 = \zeta^2 E_o r_s^2, \tag{5.2}$$

To calculate the mass we must calculate first the energy density of the electromagnetic field:

$$\rho_\varepsilon = \frac{1}{8\pi}\left(\vec{E}^2 + \vec{H}^2\right), \tag{5.3}$$

In **linear approximation** we have $|\vec{E}| = |\vec{H}|$ in Gauss's system. Then (5.3) can be written so:

$$\rho_\varepsilon = \frac{1}{4\pi}E^2, \tag{5.4}$$

Using (4.4) and a well-known relativistic relationship between a mass and energy densities:

$$\rho_m = \frac{1}{c^2}\rho_\varepsilon, \tag{5.5}$$

we obtain:

$$\rho_m = \frac{1}{4\pi\ c^2} E^2 = \frac{1}{4\pi\ c^2} E_o \cos^2 k_s l, \tag{5.6}$$

Using (5.6), we can write for the semi-photon mass (fig.1):

$$m_s = \iint_{S_t,\ l} \rho_m ds\ dl = \frac{S_c E_o^2}{\pi\ c^2} \int_0^{\frac{\lambda_s}{4}} \cos^2 k_s l\ kl, \tag{5.7}$$

From (5.7) we obtain:

$$m_s = \frac{E_o S_c}{4\omega_s c} = \frac{\pi\ \zeta^2 E_o^2 r_s^2}{4\omega_s c}, \tag{5.8}$$

### 5.2. Electromagnetic constant

Using Eqs. (5.2) and (5.8) we can write:

$$m_s = \frac{\pi\ q_s^2}{4\zeta^2 \omega_s c r_s^2}, \tag{5.9}$$

or, taking in account that $\omega_s \cdot r_s = c$ we obtain:

$$r_s = \frac{\pi}{2\zeta^2} \frac{q_s^2}{2m_s c^2}, \tag{5.10}$$

Putting here the values $\omega_s$ and $r_s$, we will obtain the value:

$$\alpha_s = \frac{q_s^2}{\hbar c} = \frac{2}{\pi} \zeta^2, \tag{5.11}$$

which corresponds to the known electromagnetic constant $\alpha = \frac{e^2}{\hbar c} \cong \frac{1}{137} \approx 0{,}007$.

The formula (5.11) shows that *in the curvilinear field theory the electric charge is defined only by the universe constants; that means that there are no free charges less than this one.*

At the same time, *our theory doesn't limit the mass value of twirled semi-photon; this fact is also according to the experimental data.*

But if $\zeta = 1$, we obtain from (5.11) $\alpha_s \cong 0{,}637$ and $q_s = 9{,}34 \cdot e$. In the following section we will try to find out, whether there is an essential reason that the charge and the fine structure constant, received according to the electromagnetic representation of the Dirac electron theory, are essentially bigger than the values known from the experiment.

### 5.3. Vacuum polarization

The experiments have shown, that the characteristics of the elementary particles are influenced by some medium, named physical vacuum. The effect of the interference of a charge and the physical vacuum is known in quantum electrodynamics, as the polarization of vacuum. In QED the physical vacuum represents, actually, an almost infinite number of enclosed in each other specific vacuums, formed by the various sorts of the virtual particles. For the dielectric permeability of such "dielectric" it is possible conditionally to write:

$$\varepsilon_\upsilon(0) = \varepsilon_\upsilon(\breve{\gamma}) + \varepsilon_\upsilon(\breve{\lambda}) + \varepsilon_\upsilon(\breve{\mu}) + \varepsilon_\upsilon(\breve{\beta}), \qquad (5.12)$$

where $\varepsilon_\upsilon(0)$ is a full dielectric constant of the physical vacuum (do not confuse with a constant $\varepsilon_0$ of the SI units system), and $\varepsilon_\upsilon(\breve{\gamma})$, $\varepsilon_\upsilon(\breve{\lambda})$, $\varepsilon_\upsilon(\breve{\mu})$, $\varepsilon_\upsilon(\breve{\beta})$ are the dielectric constants of the vacuums of the virtual photons, leptons, mesons and barions accordingly.

The polarization of the physical vacuum is considered in a number of modern reviews [15,16,17]. We will state briefly the substantive results of this theory, which are useful for our analysis.

The Coulomb potential energy [16] of systems of two charges $q'$ and $q''$ in a dielectric is equal to:

$$W(r) = \frac{q'q''}{\varepsilon_d r}, \qquad (5.13)$$

where $\varepsilon_d$ is a dielectric constant. In classical physics was accepted, that the vacuum cannot be polarized and consequently $\varepsilon_\upsilon = 1$. However, in the quantum field theory we can see something else.

The Coulomb potential energy of two charges in physical vacuum is equal to

$$W(r) = \frac{q'_0 q''_0}{\varepsilon_\upsilon r}, \qquad (5.14)$$

If to define charges $q'$ and $q''$ as follows:

$$q' = \frac{q'_0}{\sqrt{\varepsilon_\upsilon}}, \quad q'' = \frac{q''_0}{\sqrt{\varepsilon_\upsilon}}, \qquad (5.15)$$

the Coulomb energy in vacuum will take the classical form:,

$$W(r) = \frac{q'q''}{r}, \qquad (5.16)$$

The calculation in QED leads to the conclusion, that the dielectric permeability of vacuum is an infinitely large number.

It is also necessary to note [17], that the electric charge observable in the experiments, is not equal to $q_0$, but to the size $q$. The charge $e_0$, considered without the influence of polarization, is refered in QED as "bare".

Since in QED a "bare" charge is equal to infinity:

$$q^{QED}_{bare} = \infty_1, \qquad (5.17)$$

the screening charge is also equal to infinity:

$$q^{QED}_{scr} = \infty_2, \qquad (5.18)$$

The measured charge can be considered as a difference among the "bare" charge and the screening charge:

$$q_{\exp} = q^{QED}_{bare} - q^{QED}_{scr} = \infty_1 - \infty_2 = const = e, \qquad (5.19)$$

The calculation (5.19) is named in the QED as the renormalization procedure.

Thus, if it could be possible to measure an electron charge in very small distances, it would be found that in the process of penetration behind the screening layer, this charge increases. The direct consequence is the conclusion, that the electromagnetic constant is not actually a constant but it grows with reduction of distance between the cooperating particles. The known measured constant is observed on atomic distances about $10^{-8}$ cm.

Let us apply these representations to our theory. Obviously, the above calculated electron model characteristics should be considered as "bare", i.e. $q_s \equiv q_{bare}$, $\alpha_s \equiv \alpha_{bare}$, $r_s \equiv r_{bare}$, $m_s \equiv m_{bare}$.

In our theory the "bare" charge has a final size. Obviously, the screening charge also has a final size:

$$q_{bare} = q_1, \quad q_{scr} = q_2, \qquad (5.20)$$

Then the experimental charge should be defined by their difference:

$$q_{exp} = q_1 - q_2 = e, \qquad (5.21)$$

In this case for the fine structure constants the following parity takes place:

$$\alpha_{bare} \gg \alpha_{exp}, \qquad (5.22)$$

Thus, *the renormalization procedure means the account of the polarisation of the physical vacuum, when we transit from the "bare" characteristics to the experimental characteristics.*

We can try to calculate the value of dielectric permeability of vacuum at which the theoretical ("bare") value of a charge becomes equal to its experimental value. According to the above-stated formulas we can write:

$$q_{exp} = \frac{q_{bare}}{\sqrt{\varepsilon_v}}, \qquad (5.23)$$

And, according to the quantum theory:

$$q_{exp} = e = \sqrt{\alpha \, \hbar c}, \qquad (5.24)$$

Let's compare (5.24) with (5.11). Using (5.23) we receive for the fine structure constant:

$$\alpha = \frac{\alpha_s}{\varepsilon_v} \equiv \frac{\alpha_{bare}}{\varepsilon_v}, \qquad (5.25)$$

Since according to (5.11) the "bare" fine structure constant is equal to:

$$\alpha_{bare} = \frac{2}{\pi} \cong 0{,}637, \qquad (5.26)$$

the dielectric constant of the vacuum has the following value:,

$$\varepsilon_v = \frac{\alpha_q}{\alpha} \approx 87{,}27, \qquad (5.27)$$

In this case the size of a "bare" charge is equal to:

$$q_{bare} = e \cdot \sqrt{\varepsilon_o} = 9{,}34 \, e, \qquad (5.28)$$

Using our model, we can understand another one result known from the quantum mechanics. Using the model characteristics, we have received:

$$r_{bare} = \frac{1}{\alpha_q} \frac{q_{bare}^2}{m \, c^2}, \qquad (5.29)$$

Passing to the experimental charge value according to (5.28), and taking into account (5.25), we will receive:

$$r_{bare} = \frac{1}{\alpha} r_0 = r_C, \qquad (5.30)$$

where $r_0 = \dfrac{e^2}{mc^2}$ is the classical electron radius and $r_C = \dfrac{\hbar}{mc}$ is the Compton wave length of the electron. From here

$$\frac{r_0}{r_{bare}} = \alpha \approx \frac{1}{137}, \qquad (5.31)$$

Because of the incompleteness of our model, it would be naive to believe, that the found numbers are exact. Nevertheless, our results are fully concurrent to the representations of quantum electrodynamics.

### 5.4. Spin of model

Using the data of the torus model we can calculate the spin of the twirled semi-photon:

$$\sigma_s = p_s \cdot r_s = \frac{1}{2}\hbar, \qquad (5.32)$$

### 5.5. Magnetic moment of model

Magnetic moment accordingly with definition is:

$$\mu_s = I \cdot S_I, \qquad (5.33)$$

where $I$ is electron ring current and $S_I$ is the current ring square.
In our case we have:

$$I = q_s \frac{\omega_s}{2\pi} = q_s \frac{1}{2\pi} \frac{2m_e c^2}{\hbar}, \qquad (5.34)$$

$$S_I = \pi\, r_s^{\,2} = \pi \left(\frac{\hbar}{2m_e c}\right)^2, \qquad (5.35)$$

Using these formulae, we find:

$$\mu_s = \frac{1}{2}\frac{q\hbar}{2m_e}, \qquad (5.36)$$

If we put $q_s = e$, the value (3.32) is equal to half of the experimental value of the magnetic momentum of the electron. Taking into account the Thomas's precession [18] we obtain the experimental value of the electron magnetic momentum.

## 6.0. Pauli exclusion principle

The Pauli exclusion principle can be written in following form: particles of half-integer spin must have antisymmetric wavefunctions, and particles of integer spin must have symmetric wavefunctions.

As it is not difficult to see, *the Pauli exclusion principle is illustrated very well by the fig. 7, where the geometrical view of both wave - symmetric for the linear and twirled photon and antisymmetric for the electron and positron, are presented.*

The Pauli exclusion principle can also be formulated in a more concrete form: no two fermions (e.g. two electrons or positrons) can exist in identical energy quantum states. The physical origin of this fact becomes apparent from figs. 8a (two electrons) and 8b (two positrons): as we can see the electromagnetic fields of the particles do not allow them to fuse.

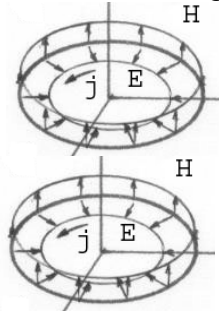
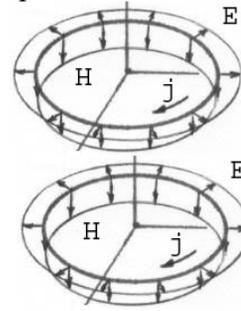

Fig. 8a          Fig. 8b

## Conclusion

Let us enumerate some results of our theory, which explain the features of the modern quantum theory:

In the papers [1,2] we received the simple and convincing explanations to all features of Dirac's equations, for example:

Why the bispinor equation contains four equations?
Why the Pauli matrices in the classical theory describe the vector rotations in the space?
Why these matrices have the certain electrodynamics vector properties and why do 16 matrices exist?
Why the Dirac's equations are received from Klein-Gordon's equation?
Why the theory of groups (i.e. the theory of the space rotation and other transformations of symmetry) is the basis for the search of the physical theory invariants?
The sense of the equivalence of Dirac's equation forms, as the equations of the waves of various space directions and twirling.
Etc.

Additionally the present paper helps us to understand many particularities of the elementary particles theory:

**The origin of the rest mass of particles**. The masses of the particles arise as the "stopped" electromagnetic energy of a twirled photon. Only the linear photon rest mass is equal to zero; all other particles should have a rest mass.
**The electric charge appearance**: it appears as the consequence of the occurrence of a tangential displacement current of Maxwell at a twirling of a photon.

**The value of the electron (positron) charge**: according to our theory it is defined only by universal constants.
**Universality of the charge**: according to our theory the charge is defined only by means of the twirling of the photons, which is the same for all Universe.
**The Pauli exclusion principle** physical origin is the symmetry or antisymmetry of the particle fields.
**The ambiguity of a charge** corresponds to the interaction of circular currents (by observance of Pauli's principle)
**The law of the electric charge conservation** follows from the way of charge formation by the division of the whole twirled photon into two twirled half periods: electron and positron.
**The neutrality of the Universe** follows also from above: the quantities of "positive" and "negative" half-periods are equal in the Universe.
**Infinite problems**: in the above theory the charge and mass infinite problems don't exist, but this fact don't contradict to the QFT.
**The spin of particles** arises owing to the twirling of the field of an electromagnetic wave.
**Helicity** is defined by poloidal rotation of a field of particles.
**The difference of the bosons and fermions**: bosons are the twirled photons and fermions are the twirled semi-photons. As other opportunities of division of a wave do not exist, there is not any other type of particles in the nature.
**The existence of particles and antiparticles** corresponds to the antisymmetry of the twirled semi-photons.
**The spontaneous breakdown of symmetry of vacuum and the occurrence of the particle mass** is connected with the change of symmetry of a linear photon at the moment of its twisting and division into two half-periods. As Higgs field here work the electromagnetic field of the nucleus.
**The Zitertbewegung** corresponds to the rotation of the semi-photon fields.
**A principle of uncertainty**: as it is known, the uncertainty relation arises in the theory of any wave. In our theory the particles are represented by the twirled electromagnetic waves. Therefore this principle is fair for elementary particles.
**An operational method**: the operators in quantum physics arise as operators of the equations of twirled electromagnetic waves theory.
**Wave packet stability of the particles**: wave packet dispersion is absent in the twirled waves thanks to their monochromaticy and a harmonicity.
**Statistical interpretation of wave function**: if an electron represents an twirled EM wave, the square of $\psi$-function will describe the density of energy of the EM field, referred to rest energy of the electron.
**The phase and gauge invariance** plays a basic role in the modern theory of elementary particles. As it is known gauge invariance means physically the same, as the phase invariance. In case that the elementary particles are the twirled electromagnetic waves, the phase invariance represents the construction tool of the theory of elementary particles.
**The wave - particle duality** ceases be a riddle in the theory of the twirled photons: elementary particles really represent simultaneously both waves and particles.
Etc.